\documentclass{aa} 
\usepackage{graphicx}
\usepackage{amsmath}
\usepackage{amsfonts}
\usepackage{amssymb}
\usepackage{gensymb}
\usepackage{textcomp}
\usepackage[varg]{txfonts}
\usepackage[T1]{fontenc}
\usepackage{epstopdf}
\usepackage{natbib}
\usepackage{hyperref}
\usepackage{placeins}
\usepackage{ulem}
\usepackage{subcaption}
\bibpunct{(}{)}{;}{a}{}{,}
\makeatletter
\renewcommand*\aa@pageof{, page \thepage{} of \pageref*{LastPage}}
\makeatother

\newcommand{\hri}{HRI$_{\rm EUV}$}

\begin{document}

\title{Stereoscopic observations reveal coherent morphology and evolution of solar coronal loops}

\author{
B.~Ram\inst{1},
L.~P.~Chitta\inst{1},
S.~Mandal\inst{1},
H.~Peter\inst{1,2},
F.~Plaschke\inst{3}
}

\institute{
Max-Planck-Institut f\"ur Sonnensystemforschung, Justus-von-Liebig-Weg 3, 37077 G\"ottingen, Germany\\
\email{ram@mps.mpg.de} 
\and
Institut für Sonnenphysik (KIS), Georges-K\"ohler-Allee 401a, 79110 Freiburg, Germany
\and
Institut f\"ur Geophysik und extraterrestrische Physik, Technische Universit\"at Braunschweig, Mendelssohnstrasse 3, 38106 Braunschweig, Germany
}

 \date{Received 23 September 2024 / Accepted 24 November 2024}
\abstract
{Coronal loops generally trace magnetic lines of force in the upper solar atmosphere. Understanding the loop morphology and its temporal evolution has implications for coronal heating models that rely on plasma heating due to reconnection at current sheets. Simultaneous observations of coronal loops from multiple vantage points are best suited for this purpose. Here we report a stereoscopic analysis of coronal loops in an active region based on observations from the Extreme Ultraviolet Imager on board the Solar Orbiter Spacecraft and the Atmospheric Imaging Assembly on board the Solar Dynamics Observatory. Our stereoscopic analysis reveals that coronal loops have nearly circular cross-sectional widths and that they exhibit temporally coherent intensity variations along their lengths on timescales of around 30\,minutes. The results suggest that coronal loops can be best represented as three-dimensional monolithic or coherent plasma bundles that outline magnetic field lines. Therefore, at least on the scales resolved by Solar Orbiter, it is unlikely that coronal loops are manifestations of emission from the randomly aligned wrinkles in two-dimensional plasma sheets along the line of sight, as proposed in the coronal veil hypothesis.}

\titlerunning{Coherent morphology and evolution of solar coronal loops}
\authorrunning{B.~Ram et al.}

 \maketitle

\section{Introduction}
Coronal loops are the bright arched structures observed in the extreme ultraviolet (EUV) and X-ray wavelengths. 
They are associated with closed magnetic flux tubes that confine hot plasma, with temperatures exceeding one million Kelvin \citep{Aschwanden_2002_review, Reale_2014_review}. The mechanisms responsible for such intense heating are a topic of ongoing debate. There are two widely studied mechanisms to explain the loop heating. Several studies suggest that magnetohydrodynamic (MHD) waves that dissipate their energy at the coronal heights are responsible for loop heating \citep[e.g.,][]{Hollweg_1981, Walsh_2003, Asgari_2015, Doorsselaere_2020}. However, such wave-heating scenarios seem to be insufficient to sustain the high coronal temperatures found in active regions (ARs) \citep{Withbroe_1977, McIntosh_2011}.
As an alternative to wave-heating, other studies propose that numerous small-scale bursts of energy, termed nanoflares, impulsively heat coronal loops to the observed temperatures \citep[e.g.,][]{Parker_1988,
Cargill_2004, Klimchuk_2006}. In reality, both of these mechanisms (wave-heating and nanoflares) may be operating, although at different spatial or temporal scales. Studying the loop morphology and evolution will then provide additional constraints on coronal heating models. 

It has been suggested that coronal loops are composed of strands with thicknesses approaching the spatial resolution limit of the observing instrument \citep{Matteo_1999, Testa_2002}. This basically implies that the dominant heating mechanism may be operating on very small spatial scales, where each strand is rapidly heated by storms of nanoflares \citep{Warren_2003}. Based on the visibility of X-ray loops for durations longer than the typical cooling timescales, \cite{Fuentes_2007} suggested that loop-averaged heating rates due to nanoflares first slowly increase, come to a maintenance level, and then decrease slowly. Based on these observations, 1D simulations by \cite{fuentes_2010} demonstrate that energy deposition in the form of nanoflares can rapidly heat the entire loop, leading to an almost instantaneous brightness enhancement in passbands sensitive to emission from hotter plasma. As the loops cool, they progressively appear in channels sensitive to successively cooler temperatures, which explains the observed time lag in the peak brightness among different channels \citep{Warren_2003, Winebarger_2003, Winebarger_and_Warren_2005, Viall_2012}. These findings imply that during the cooling phase, the intensity along the entire length of a loop should evolve similarly, even if individual strands are heated independently.
However, the three-dimensional (3D) structure of these loops is crucial for identifying where and how the energy is deposited to heat the plasma. This hinders our understanding of the mechanisms responsible for their heating to high temperatures of over a million kelvin.

X-ray and EUV observations generally suggest that the coronal loops have uniform thicknesses along their length. This was first reported by \cite{Klimchuk_1992} for soft X-ray and EUV loops. Later, \cite{Watko_and_Klimchuk_2000} utilized Transition Region and Coronal Explorer \citep[TRACE;][]{Handy_1999} observations to examine width variations along coronal loops, and found that widths remained largely consistent from footpoint to apex.
\cite{Klimchuk_2020} analyzed 20 High-Resolution Coronal Imager \citep[Hi-C;][]{Kobayashi_2014} loops and observed no significant negative correlation between loop intensity and width. These findings suggest a circular cross section for the loops assuming it has a nonnegligible twist and roughly uniform plasma emissivity along the magnetic field.
Additionally, \cite{Kucera_2019} compared the widths of two loops and their line-of-sight (LOS) thicknesses deduced using spectroscopic observations from the Hinode mission \citep{Culhane_2007, Kosugi_2007}. Assuming a filling factor close to unity, the findings also suggest that the observations align with a circular cross section.
Recently, \cite{Mandal_2024} observed a loop from two vantage points using data obtained from the EUV High-Resolution Imager (\hri), part of the Extreme Ultraviolet Imager \citep[EUI;][]{Rochus_2020} on board Solar Orbiter \citep[][]{2020A&A...642A...1M}, and data from the Atmospheric Imaging Assembly on board the Solar Dynamics Observatory \citep [SDO/AIA;][]{Pesnell_2012, Lemen_2012}. They also concluded that the observed loop maintains a circular cross section since it showcases similar widths from both vantage points and along its length.

As the magnetic field expands with height and becomes space-filling in the corona, conventional magnetic field extrapolation models generally predict coronal loops expanding with height. 
Several explanations have been proposed for this lack of expansion in observations.
\cite{Klimchuk_2000} demonstrated that a locally enhanced twist in the field lines can force a loop to be circular and decrease the expansion. However, the required amount of twist might make loops kink unstable \citep{Amari_1999}.
\cite{Plowman_2009} proposed that loops that form close to magnetic separators would show a constant width along their length. However, it is uncertain what fraction of the loop population is associated with separators. 
\cite{Peter_and_Bingert_2012} suggested that loops might have a temperature gradient across their lengths that makes only a part of their lateral extent visible in a particular wavelength band. Nonetheless, the mechanism responsible for generating this systematic cross-field thermal structure is unclear \citep{Klimchuk_2020}.

On the other hand, some studies suggest that loops expand with height and have elliptical cross sections. \cite{Wang_and_Sakurai_1998} found a small anisotropic expansion in highly sheared loops in their axisymmetric nonlinear force-free field models.
Using 3D numerical MHD simulations, \cite{Gudiksen_2005} also showed that flux tubes generally do not expand identically in all directions with height. 
Additionally, \cite{Fuentes_2006} observed that TRACE loops were more symmetric than the corresponding flux tubes extrapolated using linear force-free field extrapolation models.
Furthermore, \cite{Malanushenko_2013} highlight a potential selection bias, where observers might favor loops primarily expanding in the LOS direction, leading to the misinterpretation of a non-expanding loop in the plane of sky as circular in cross section.

Recently, using 3D MHD simulations of the solar corona including magnetoconvection, \cite{Malanushenko_2022} demonstrated that integrated emission from thin veil-like sheets in the corona can also be confused as coronal loops. The authors argue that this hypothesis can explain the constant cross sections and anomalously high-density scale-height in the corona. Since veil-like sheets represent a broader and higher-entropy range of shapes compared to isolated compact strands, and observations have not definitively ruled out their presence, they suggest that coronal veils should be considered the most plausible scenario when interpreting observations of coronal loops.

For this paper we studied the morphology and evolution of coronal loops from two vantage points by utilizing the data sets obtained from the \hri\ and SDO/AIA. This gives us a unique opportunity to conduct high-resolution stereoscopic observations to further investigate the cross-sectional structure of coronal loops and their evolution along their lengths. Identifying the same loops in the two instruments and comparing their widths and thermal evolution would provide important insights into the 3D structure of coronal loops and their underlying heating mechanisms.

\begin{figure}
 \centering
 \includegraphics[width=\linewidth]{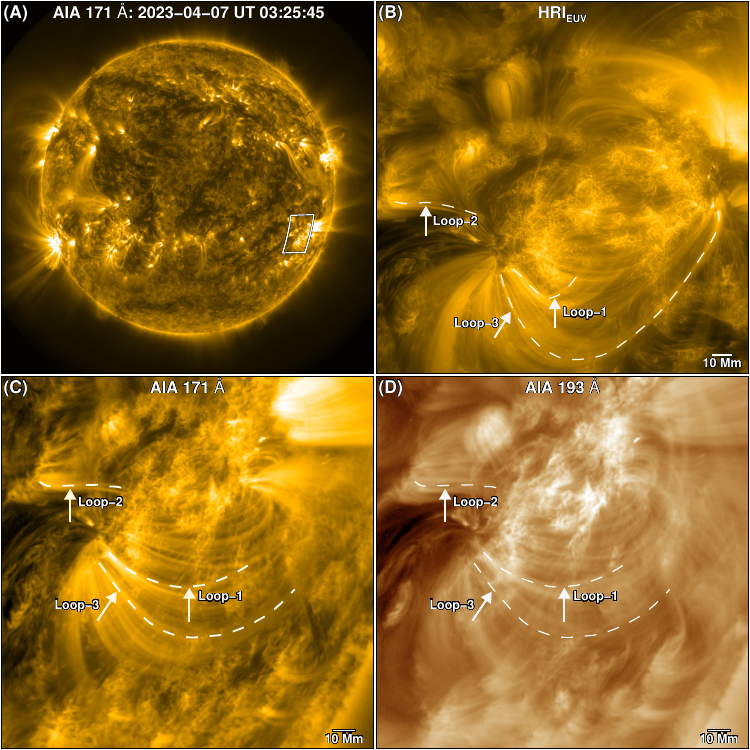}
 \caption{Overview of the observations. Panel (A): Full disk image of the solar corona as observed with 171~{\AA} filter of the SDO/AIA. Panel (B): Part of the field of view of the \hri\ that shows active region loops. This field of view is outlined by the white box in panel (A). Panels (C) and (D): Roughly the same region as seen by the AIA 171~{\AA} and 193~{\AA} filters, respectively. The white dotted lines trace the three loops described in the text. The intensity in panel (A) is displayed in square root scale. Panels (B)-(D) are time averages of three images for each passband that are closest in time with the image shown in panel (A) and are plotted in log scale. }
 \label{ref1}

\end{figure}

\section{Observations and data processing}

\begin{figure*}
 \centering
 \includegraphics[width=\textwidth]{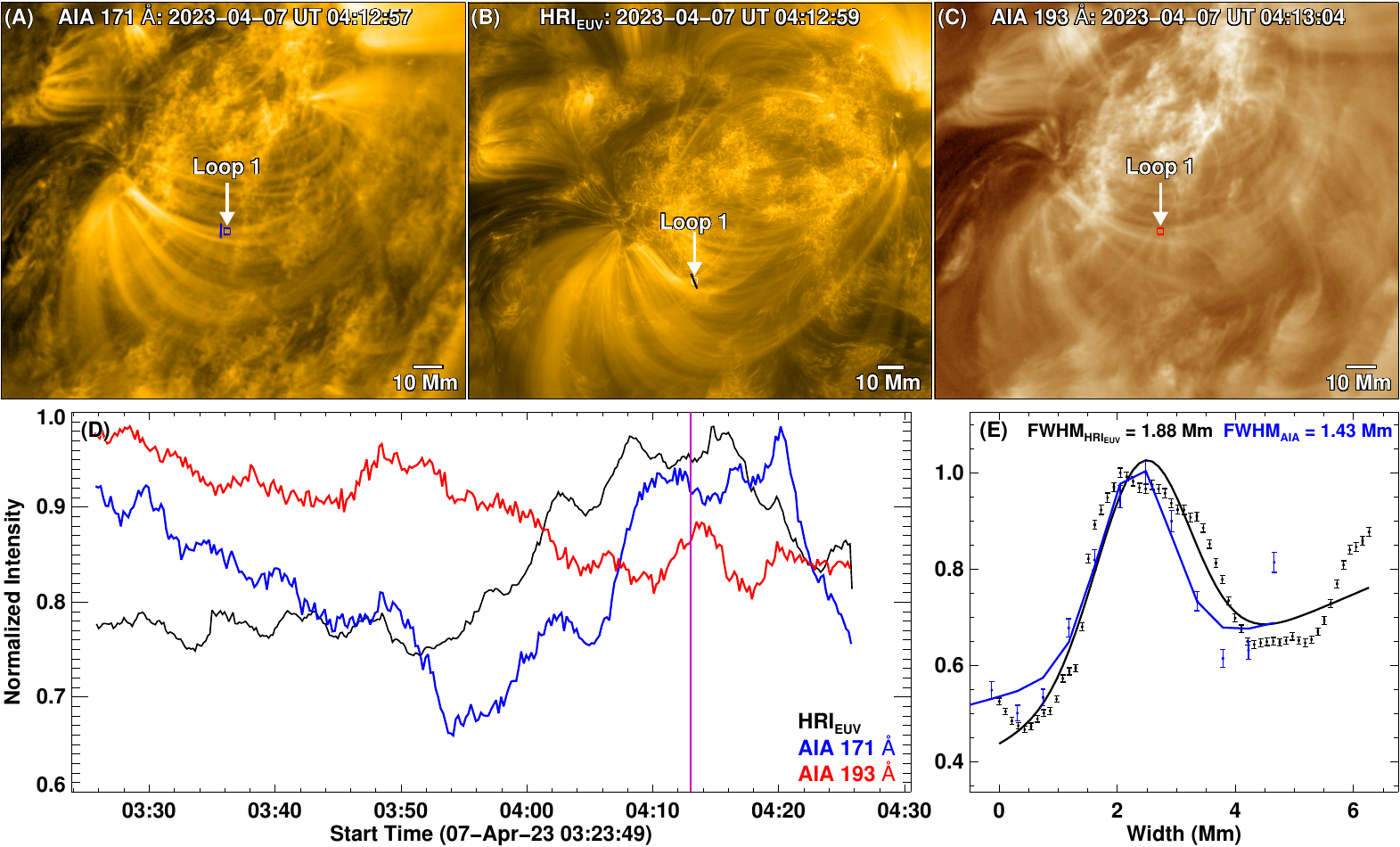}
 \caption{Evolution of loop 1. Panels (A)-(C) show the snapshots of loop 1 in the AIA 171~{\AA}, \hri, and AIA 193~{\AA} passbands, respectively. The arrows highlight the color-coded boxes spanning 5~$\times$~5 pixel in both of the AIA passbands and 20~$\times$~20 pixel in the EUI 174~{\AA} passband. The average intensities within these boxes are utilized to compute the light curves, which are subsequently normalized by their individual maximum values. These normalized light curves are plotted in panel (D) using the same colors as their corresponding boxes. The color-coded solid lines near the boxes in panels (A) and (B) represent roughly 1.3 Mm wide artificial slits that we used to calculate the loop widths in panel (E). The lengths of these slits differ for different loops depending upon the loop width and the presence of adjacent loops. The data points in panel (E) correspond to the average intensities observed along the artificial transverse slits, and the solid lines show their respective Gaussian fits. The intensities in panel (E) are normalized with the respective maxima. The error bars show 1{$\sigma$} uncertainties in the measurements due to the Poisson errors that arise from photon noise. Panels (A)-(C) are plotted in log scale. An animation of panels (A)-(D) is available \href{https://www.dropbox.com/scl/fi/w25x84sorew7s3todv6ie/32.mp4?rlkey=fdvyz3evkzavqztw3xdwtorke&dl=0}{online}. }
 \label{loop1}
\end{figure*}

We analyzed the data obtained from the EUI on board the Solar Orbiter spacecraft and the AIA on board the SDO. During this period, the field of view of the \hri\ includes the AR NOA13270 located close to the limb in the southern hemisphere (see Fig. \ref{ref1}A).
The 174~{\AA} \hri\ of the EUI captured this data set on April 7, 2023 \citep[][]{euidatarelease6}, when the Solar Orbiter spacecraft was at a distance of $\sim$0.3 astronomical units (au) from the sun. The \hri\ has a plate scale of 0.492\arcsec\,pixel$^{-1}$, which corresponds to a pixel size of roughly 108\,km at this distance. The core of the \hri\ point spread function (PSF) has a full width at half maximum (FWHM) of $\sim$2 pixel, resulting in a spatial resolution of about 216\,km. These observations start from 03:25:49 to 06:50:40 UT with a variable cadence that shifts from 10~s to 3~s and back to 10~s. We removed the jitter in the level 2 data by applying the cross-correlation technique described in \citet{Chitta_2022}.

The corresponding AIA data were respiked using the SolarSoft procedure aia\_respike. We used aia\_prep to eliminate the spacecraft roll angle, update the pointing information, and set the plate scale to 0.6$\arcsec$ per pixel. Additionally, we registered the entire AIA data set with the first AIA image to remove the solar rotation using coreg\_map.
We utilized a subset of the \hri\ data set with a cadence of 10~s spanning from 03:25:49 to 04:25:49 UT because it closely matches the 12~s cadence of the SDO/AIA filters. 

To align the AIA data with the \hri\ images, we extracted World Coordinate System \citep[WCS;][]{Thompson_2006} information from the EUI header files and calculated the Carrington coordinates using the IDL routine wcs\_convert\_from\_coord. Using wcs\_convert\_to\_coord, we then calculated the corresponding world coordinates in AIA passbands. To refine alignment, manual adjustments were made by tracking common features in the \hri\ images and the AIA 171~{\AA} passband.

Figure \ref{ref1} provides a comprehensive view of the region of interest (ROI). We focus on the bright closed loops with relatively low backgrounds in the AR. At the time of observation, the angle between the Solar Orbiter and the Sun--Earth line is $\sim$43$^{\circ}$, resulting in the foreshortening of these loops in the AIA images. Despite this distortion, we were able to visually identify ten prominent loops consistent across the two instruments, based on their footpoints and similar morphology.
\begin{figure*}
 \centering
 \includegraphics[width=\textwidth]{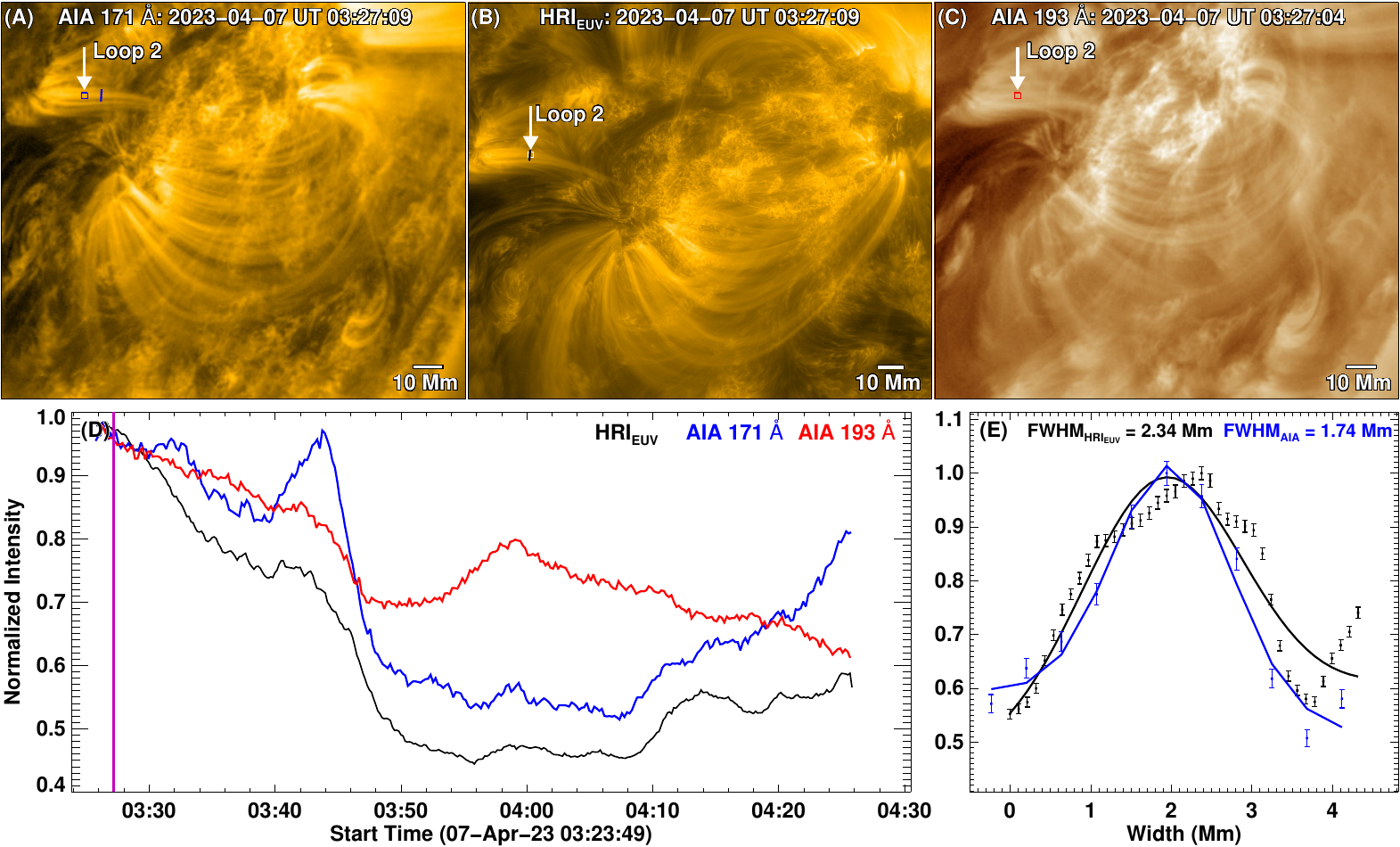}
 \caption{Evolution of loop 2. The panel layout, box sizes for light curves, and slit width are the same as for Fig. \ref{loop1}. An animation of panels (A)-(D) is available \href{https://www.dropbox.com/scl/fi/mki4qfge0p6bxpg7z94wu/26.mp4?rlkey=sdmlohp5nvt3gck5tj4qq1h7k&dl=0}{online}.}
 \label{loop2}
\end{figure*}

 The \hri\ instrument on board Solar Orbiter samples the plasma at $\sim$1\,MK; Fe\,{\sc ix} and Fe\,{\sc x} are the dominant species that contribute to the emission detected by the passband. The AIA 171~{\AA} filter captures the plasma at $\sim$0.8~MK primarily dominated by Fe IX ions. This overlap in temperature response functions coupled with the inherent multi-thermal structure of the coronal loops causes them to exhibit similar temporal evolutions when observed with these two instruments.
 This similarity aids in identifying the same loops in data sets obtained from two vantage points making them ideal for stereoscopic studies of coronal loops. Additionally, the AIA 193~{\AA} channel is dominated by Fe\,{\sc xii} ions in quiet-Sun and nonflaring ARs, and samples the plasma at $\sim$1.5~MK. This difference in temperature response between the AIA 193~{\AA}, and the cooler AIA 171~{\AA} filter and the \hri\ offers possibility to investigate multi-thermal evolution of loops, providing a more comprehensive view of their dynamics.

 \section{Results}

\subsection{Similar temporal evolution of loops from two vantage points}

Following the visual identification of each loop, we then compared their temporal evolution across both instruments using light curves. 
These light curves were generated by averaging the intensity within $\sim2.2\times2.2$~Mm boxes (20$\times$20 pixel in \hri\ and 5$\times$5 pixel in AIA) positioned on the loop axes. Notably, these boxes are not remapped between the \hri\ and AIA instruments. 
We manually position the boxes in both data sets on loop segments with darker backgrounds.
The \hri\ box position was fixed initially, while the box in AIA 171~{\AA} filter was incrementally moved along the loop within the low-background segment. If the light curves from these boxes displayed similar intensity variations, then the \hri\ box position was slightly shifted along the loop length and the process was repeated. This ensured that the temporal evolution of the entire \hri\ loop segment with a low background matched the corresponding segment in the AIA 171~{\AA} filter.

We also compared the light curves of neighboring loops in the AIA 171~{\AA} passband with the identified loop in the \hri\ data.
We found that adjacent loops exhibited noticeably different temporal variations on longer timescales (see Appendix \ref{adjacent_loop_appendix}). Therefore, it further ensured that we are identifying the same loop in the AIA 171~{\AA} filter and are not confusing it with an adjacent loop.
The shorter observation period, which spans $\sim$1 hr, minimizes significant movements of the loops in the plane of the sky. We also reviewed the accompanying movies to ensure that the light curves are unaffected by loops moving out of the boxes. Though rare, if this movement occurs, the box placement is adjusted accordingly. We performed this analysis on ten visually identified loops. Three such loops are highlighted in panels (B)-(D) of Fig. \ref{ref1}.\\

\begin{figure*}
 \centering
 \includegraphics[width=\linewidth]{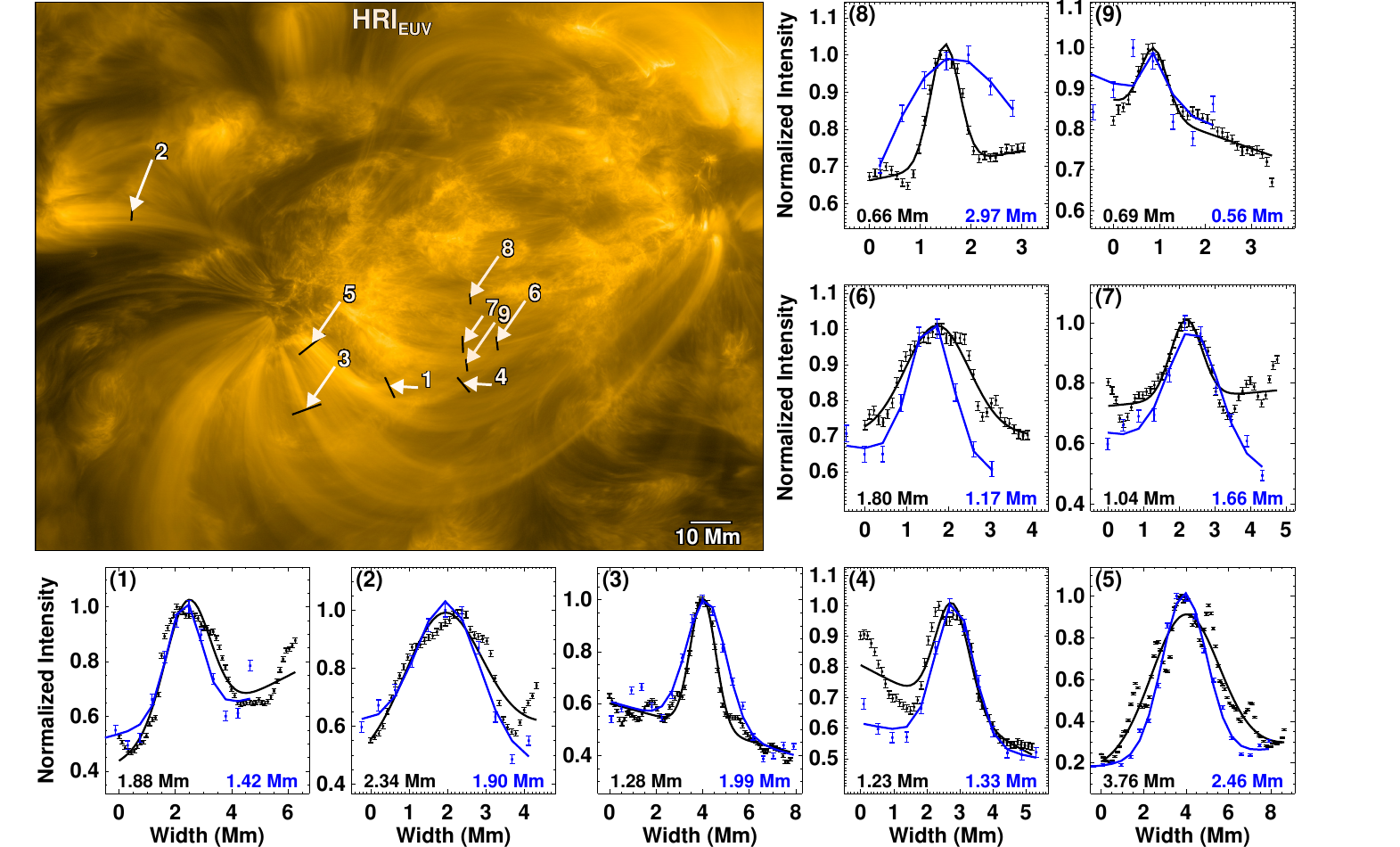}
 \caption{Comparison of loop widths from two vantage points. For visualization purposes, we selected a single \hri\ snapshot for each loop at its peak brightness. The top left panel displays the average of these nine snapshots. The black lines indicate slit positions used for width calculations in the \hri\ data. Panels (1)-(9) plot the intensity profiles for the individual loops (numbered 1-9) corresponding to the positions in the top left panel. Measurements in black (\hri) and blue (AIA 171~{\AA}) represent the loop widths as described in the caption of Fig. \ref{loop1}E. These loop widths are tabulated in Appendix \ref{table_1}}
 \label{loop_widths}
\end{figure*}

Figure \ref{loop1} illustrates the evolution of loop 1 located in the core of the AR. 
The accompanied animation shows that the loop is distinctly visible in the AIA 171~{\AA} filter and \hri\ images, with a faint presence in the AIA 193~{\AA} passband at the beginning of the observations. Subsequently, the loop becomes fainter across all three channels, but it almost fades away in the AIA 193~{\AA} filter (03:48~-~03:56 UT, see Fig. \ref{loop1}D and accompanying animation). Later, a resurgence occurs, and the loop becomes bright in all three channels almost simultaneously in the AIA 171~{\AA} passband and \hri\ images (04:00 UT), followed slightly later by the AIA 193~{\AA} passband (04:11 UT). This brightening is transient, as the loop once again becomes faint across all three channels almost concurrently. This recurring trend is evident in panel (D) of Fig. \ref{loop1} where we show the light curves for the loop apex.

This trend is noticeable not only within the AR core, but also in the loops extending beyond it. Figure \ref{loop2} focuses on a loop consisting of multiple strands. These strands originate in the periphery of the AR core and extend to another footpoint that is not not apparent, but seems to be somewhere near the core. Panel (D) of this figure demonstrates the similar evolution of one of these strands observed in the \hri\ and the AIA 171~{\AA} passband. Initially, during the start of the observations, the strand is distinctly visible within the loop in the \hri\ and the AIA 171~{\AA} channels; however, it appears fuzzy, but still distinguishable in the AIA 193~{\AA} images. As the evolution progresses the loop fades away in the \hri\ and the AIA 171~{\AA} passband (at around 03:46 UT; see Fig. \ref{loop2}D and accompanying animation), while it becomes more distinct and luminous in the AIA 193~{\AA} channel (around 04:00 UT). By the end, it becomes distinctly visible in both the \hri\ images and the AIA 171~{\AA} filter; however, the entire loop becomes fuzzy in AIA 193~{\AA}. It suggests that this loop is also showing nearly identical evolution across the \hri\ and the AIA 171~{\AA} filter.\\

This analysis reveals that the identified loops exhibit intensity variations predominantly on two distinct timescales: a shorter scale of 5--10 minutes and a longer scale of 20--30 minutes. We find that the evolution of these loops does not appear to correlate on shorter timescales. However, we observe very similar evolution patterns between the \hri\ images and AIA 171~{\AA} passband on longer timescales. Most of these loops appear fuzzy in the AIA 193~{\AA} passband, and their light curves exhibit noticeably different evolution patterns compared to the \hri\ and AIA 171~{\AA} light curves on both shorter and longer timescales.

\subsection{Loop widths}

Furthermore, we compared the widths of these loops as observed in the \hri\ images and the AIA 171~{\AA} passband (see panel (E) of Figs. \ref{loop1} and~\ref{loop2}, and panels (1-9) of Fig. \ref{loop_widths}). To calculate the widths, we placed a $\sim$1.5~Mm wide slit across the length of the loop near the points where the loop displays a coherent evolution across these instruments. The average intensity profile (for three consecutive frames) along the slit is then fitted with a single Gaussian model with five free parameters, including a constant term and a linear trend. The FWHM derived from these fits serves as a measure of the loop cross-sectional widths. The loops in our observations exhibit widths in the range of $0.56~\pm~0.15$ Mm to $3.76~\pm~0.01$ Mm. Notably, some loops in the \hri\ images exhibit substructure, with strands appearing as multiple peaks in the intensity profiles used for these Gaussian fits. However, in most cases, the AIA 171~{\AA} counterparts lacked the spatial resolution to resolve these individual strands. For instance, loop 1 displays two strands in the \hri\ images which are evident due to a small intensity dip in the middle of the profile (black curve in panel (E) of Fig. \ref{loop1}). In contrast, the AIA 171~{\AA} curve does not resolve these individual strands, necessitating the use of a single-peaked Gaussian function for width measurement. 
Additionally, the width of loop 8 in Fig. \ref{loop_widths} was measured to be 2.97~Mm in AIA 171~{\AA}, due to the bright background moss region artificially making the loop appear thicker. Nevertheless, we identified a similarity in widths between the \hri\ images and the AIA 171~{\AA} passband, with the largest difference (excluding loop 8) being 1.3 Mm.

\subsection{Coherency}
\begin{figure*}
 \centering
 \includegraphics[width=\linewidth]{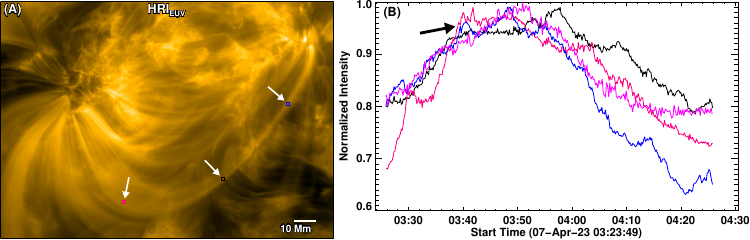}
 \caption{Coherent evolution of a loop. Panel (A) shows a one-hour time average of \hri\ images, starting at 03:23:49 UT on 07 April 2023. Three arrows point to 15~$\times$~15 pixel color-coded boxes representing different segments of loop 3 (see Fig. \ref{ref1}). Panel (B) displays colored light curves for these boxes, calculated using the process outlined in the caption of Fig. \ref{loop1}. The black arrow highlights the intensity enhancement exhibited by all the loop segments at around 03:40 UT. The image in panel (A) is plotted in log scale.}
 \label{loop_evolution}
\end{figure*}

The entire loop 3 is distinctly visible throughout the observational period in \hri\ images. This visibility facilitates a detailed comparison with other segments of the same loop that lack the underlying moss emission. Figure \ref{loop_evolution} illustrates the intensity variation for three such segments of this loop. The average intensity inside all three boxes displays a rise phase, a plateau, and then a gradual decrease. Although the intensity variations across the three segments do not correlate well over shorter timescales, there is a clear similarity in their long-term variations. This indicates a coherent thermal evolution along the length of the loop. 
Interestingly, a segment of this loop near the west footpoint does not show a clear temporal coherence between the \hri\ and AIA 171~{\AA} filter (see Appendix \ref{Dissimilar_evolution}). A prominent peak seen in the three moss-free segments around 03:40~-~04:00 UT is also faintly present in this loop segment (indicated by black arrows in Figs. \ref{loop3}D and \ref{loop_evolution}B). However, these three loop segments do not exhibit enhanced brightening toward the end of the observational period, unlike the loop segment with moss in Fig. \ref{loop3}. We also include the intensity variation of a segment of loop 3 that is outside of the moss region, as seen by the AIA 171~{\AA} filter, as a magenta curve. We find that it also displays a prominent peak at $\sim$~03:50 UT and no noticeable brightening by the end of the observation period. This is in line with the \hri\ segments that lack the moss emission in the background. This emphasizes the consistency observed in segments that are minimally affected by the background moss emission, indicating similar behavior between the \hri\ and AIA 171~{\AA} data.

\section{Discussion}

Accurately matching corresponding loop segments between different vantage points can be challenging. \cite{Malanushenko_2013} pointed out that the presence of multiple loops with elongated cross sections in close proximity can lead to false identifications, due to the potential overlap or resemblance among adjacent loops.
Our analysis reveals that the identified loops predominantly exhibit intensity variations on two distinct timescales. The short-term oscillations, in the 5--10 min range, are associated with the background moss regions and can be linked to p-mode oscillations \citep{Brynildsen_2003, Marsh_2006}. On a longer timescale of 20--30 minutes, we observe consistent intensity variations across the instruments. These longer timescale oscillations are typically associated with the cooling times of coronal loops. Consequently, if instruments observe plasma at similar temperatures, a coronal loop with a 3D structure is expected to display a similar evolution pattern from two vantage points. Notably, these observations are taken at a separation angle of approximately 43\textdegree, implying that the background and foreground features differ significantly for observations from each instrument. Nevertheless, we investigated whether the background could influence the light curves used to compare the thermal evolution of these loops across instruments. Our findings demonstrate that as long as the background is sufficiently dark, it does not substantially affect the observed intensity variations (see Appendix \ref{BG_influence_appendix}) and aligns with the results from \cite{Fuentes_2008} on TRACE loops.
Therefore, the similarity in emission patterns from two vantage points is unlikely to originate from features other than loops. Additionally, the neighboring loops show different emission patterns in the light curves. This further highlights the reliability of our analysis.

\subsection{Spatio-temporal morphology of coronal loops}
We proceeded to measure the widths of these loops as observed from both vantage points. The widths of the loops included in this study range between 0.67 and 3.76 Mm, consistent with the findings from previous studies of EUV loops \citep{Patsourakos_2007, Aschwanden_2008, Aschwanden_2017}. We observed that these loops generally have similar widths across the instruments, with the largest difference being 1.3 Mm. This suggests that these loops have nearly circular cross sections, as recently found by \cite{Mandal_2024}. The thinnest loop included in this study has a width of 660~km in the \hri\ images. Conversely, the Gaussian profile of the AIA counterpart of this loop has a width of 2.97 Mm. This discrepancy is likely attributable to the bright background that may artificially render these loops thicker. Additionally, \cite{Peter_2013} identified miniature loops in Hi-C observations as short as 1~Mm and approximately 200~km wide. However, they did not find any evidence of substructure in the larger loops at scales of 0.1\arcsec\,pixel$^{-1}$. They estimated that for these loops to be multi-stranded without showing substructure, the individual strands would need to be narrower than 15\,km. 

The consistency in intensity variations and widths of the loops observed from two different vantage points implies that the emission likely emanates from 3D structures rather than flat sheet-like structures aligned along the LOS, as suggested by \cite{Malanushenko_2022}. Our observation of a coherent thermal evolution along the entire length of an individual loop, based on stereoscopic analysis, further supports this finding. This coherence indicates that the emission originates from a 3D structure with a heating mechanism that elicits a global response across the loop. Such a structure could be composed of thin unresolved strands. Alternatively, a substructure of a loop might arise from the emission from overlapping contributions of numerous randomly oriented wrinkles within a single 2D plasma sheet. While this scenario might create an impression of a 3D structure, it would require these tiny veils to be confined within a highly unlikely circular envelope. Furthermore, even under such conditions, it is unlikely that this arrangement could consistently produce similar intensity variations when observed from two viewing angles separated by 43\textdegree, as the emission from the second perspective would come from different segments of these wrinkles. \cite{Uritsky_2024} investigated the influence of projection effects on 3D structures with varying cross sections and two-dimensional (2D) plasma sheets in an optically thin corona. They suggested that most EUV coronal loops are 3D tube-like structures and are not likely to be projections of 2D sheets. The coherency is not only imprinted in the thermal evolution as shown in this study, but also in the signatures of decayless kink oscillations of coronal loops as recently inferred using coronal observations from two vantage points \citep[][]{Zhong_2023}. 

\subsection{Nature of coronal loop heating}
The observed loops could be heated by storms of nanoflares events \citep{Klimchuk_2023}, facilitated by untangling of small-scale magnetic braids in the corona \citep{Chitta_2022}. Further observations are needed to determine the specific distribution of nanoflares within these structures. One key question is whether these nanoflare storms are localized at a single point of the loop or are distributed throughout its entire length. In either case, thermal conduction would rapidly distribute the heat, leading to a consistent intensity evolution pattern along the length. If the storms are distributed along the length of the loop, then what triggers their simultaneous occurrence remains an open question. Simulations do show that a single unstable strand experiencing kink instability can trigger an avalanche, thus disrupting neighboring strands \citep{Hood_2016}. \cite{Reid_2020} demonstrate that the kink instability heats the first strand primarily through strong bursts of Ohmic heating. As the cascade progresses, viscous heating becomes dominant, occurring in a large number of small heating events. They find that there is no clear preferred location for heating within the loop, and the heating events are spread throughout the inner 90\% of the loop length. 

With the spatio-temporal coherence of loop emission presented here, our study tends to support the conclusions from the MHD simulations. However, a more quantitative comparison of emission patterns of simulated loops (both spatial and temporal) with the type of observations we presented here is required to fully comprehend the spatial distribution of heating events in coronal loops. For instance, high-resolution MHD simulations that use self-consistent convective driving to heat coronal loop plasma \citep{Breu_2022} will be useful for such a comparison. The slow footpoint driving leads to the development of turbulence in the coronal section of the loop in these simulations. This turbulence may, on the one hand, lead to a uniform distribution of heating events, and on the other hand, smooth out the current sheets to form coherent strand-like or loop-like features. 

\subsection{Line-of-sight effects}
We note that the possibility of some coronal loops being projection effects of 2D plasma sheets cannot be completely ruled out with the current observations as the widths measured in this study may not fully account for the true cross-sectional shapes of the loops. If these loops were indeed 2D sheets, their apparent widths would vary depending on the angle between the LOS and the plane of the sheet. For example, if the LOS were parallel to the plane of a light emitting sheet, then it would appear to have its actual width, while it would be invisible if the LOS were perpendicular to the plane of the sheet. Intermediate angles cause the sheet to appear narrower, resembling loops with varied widths depending on different vantage points. To disentangle these effects, observations from a second vantage point from outside the ecliptic plane could be helpful. By viewing the structure from multiple planes, it would be possible to measure its actual width and determine its three-dimensional structure more accurately.

Practically, however, a north--south aligned coronal loop seen from different viewing angles from the same plane might also help in resolving the ambiguity between 2D sheet and 3D tube morphology. In our observations, there are indeed loop segments that are aligned nearly in the north--south orientation (e.g., the case presented in Fig.\,\ref{loop3}). The loop segment indicates more of a 3D tube-like morphology instead of a 2D sheet-like feature. Therefore, we tend to agree with the suggestion that the observed loops are 3D features rather than projections of randomly aligned wrinkles in 2D plasma sheets.

\section{Conclusions}
We analyzed AR coronal loops from two vantage points using the Solar Orbiter and Solar Dynamics Observatory, with a separation angle of $\sim$43\textdegree\ between their LOSs. This approach may provide insights into their three-dimensional geometry, including cross sections and coherent thermal evolution. 
These coronal loops appear foreshortened in the AIA images since the ROI is close to the limb. Nevertheless, we were able to identify ten distinct loops that were relatively isolated and present across both instruments, the \hri\ and in the AIA 171~{\AA} passband. The identified loops were scattered throughout the AR, with some located close to the core and others near the boundaries. After accounting for the possible contamination from the background features, our analysis suggests that the coronal loops show similar evolution of emission patterns from two different vantage points and along their lengths on timescales of about 30 minutes. The similar intensity variation across the instruments shows that we are indeed observing the same segment of a coherent 3D structure in the solar corona. We also find that these loops have similar widths across the two instruments, which suggests that they have nearly circular cross sections. 
These results hint that coronal loops are unlikely to be imprints of LOS-integrated emission from a distribution of 2D veil-like structures.

\begin{acknowledgements}
We thank the anonymous referee for their helpful comments that improved the presentation of the manuscript. B.R. and L.P.C. gratefully acknowledge funding by the European Union (ERC, ORIGIN, 101039844). Views and opinions expressed are however those of the author(s) only and do not necessarily reflect those of the European Union or the European Research Council. Neither the European Union nor the granting authority can be held responsible for them. The work of S.M. has been supported by Deutsches Zentrum für Luft- und Raumfahrt (DLR) through grant DLR\_FKZ 50OU2201. Solar Orbiter is a space mission of international collaboration between ESA and NASA, operated by ESA. We thank the ESA SOC and MOC teams for their support. The EUI instrument was built by CSL, IAS, MPS, MSSL/UCL, PMOD/WRC, ROB, LCF/IO with funding from the Belgian Federal Science Policy Office (BELSPO/PRODEX PEA 4000106864 and 4000112292); the Centre National d’Etudes Spatiales (CNES); the UK Space Agency (UKSA); the Bundesministerium für Wirtschaft und Energie (BMWi) through the Deutsches Zentrum für Luft- und Raumfahrt (DLR); and the Swiss Space Office (SSO). SDO is the first mission to be launched for NASA's Living With a Star (LWS) Program and the data supplied courtesy of the HMI and AIA consortia. This research has made use of NASA’s Astrophysics Data System Bibliographic Services.
\end{acknowledgements}

\bibliographystyle{aa}
\bibliography{ref}

%\onecolumn
\begin{appendix}

\section{Comparison of the intensity pattern of two adjacent loops}
\label{adjacent_loop_appendix}
\begin{figure}[h!]
 \centering
 \includegraphics[width=\hsize]{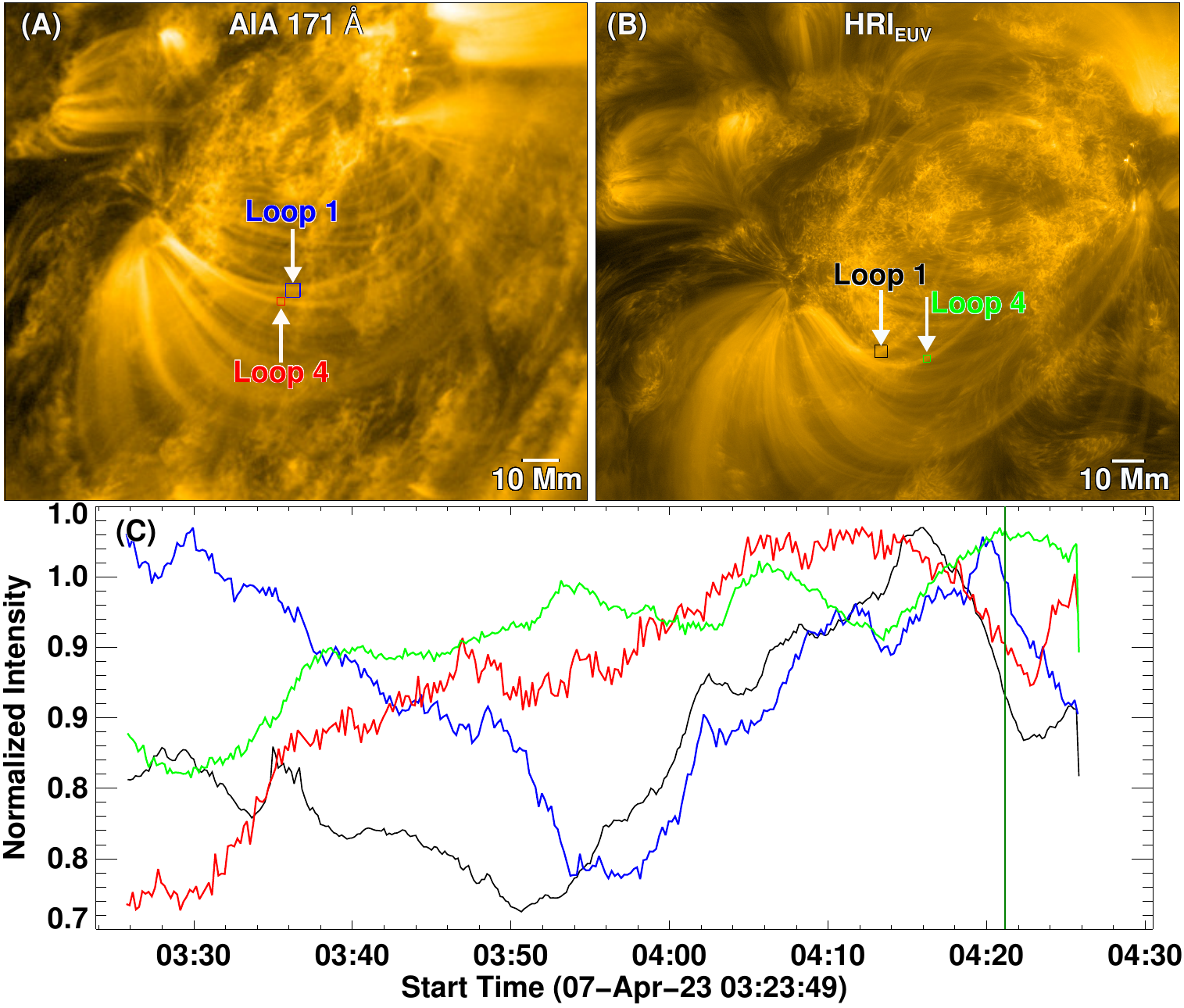}
 \caption{Difference in the evolution patterns of adjacent loops. Panels (A) and (B) showcase snapshots of the AIA 171~{\AA} and \hri\ passbands, respectively, at 04:21:09 UT. Panel (C) shows the color-coded light curves for the boxes marked in panels (A) and (B). Procedure to calculate light curve is described in the caption of Fig. \ref{loop1}. The box size for loop 1 in AIA and \hri~are increased to 9~$\times$~9 and 36~$\times$~36 pixel, respectively.}
 \label{adj_loops}
\end{figure}

The light curves from adjacent loops exhibit noticeably different emission patterns. Figure \ref{adj_loops} displays the time evolution of loop 1 and loop 4 (see also Fig. \ref{loop_widths}). Initially, the light curves from loop 1 (black and blue) show a decrease in intensity in both instruments. This intensity dip around 4:52 UT is followed by an increase, with the peak appearing slightly later in AIA 171{\AA}. This delay is likely due to the AIA 171~{\AA} filter being more sensitive to slightly cooler temperatures than \hri. In contrast, the light curves of loop 4 (green and red) exhibit a steady increase in intensity throughout the observation period.

\section{Dissimilar temporal evolution of a loop from two vantage points}
\label{Dissimilar_evolution}

\begin{figure*}
 \centering
 \includegraphics[width=0.89\linewidth]{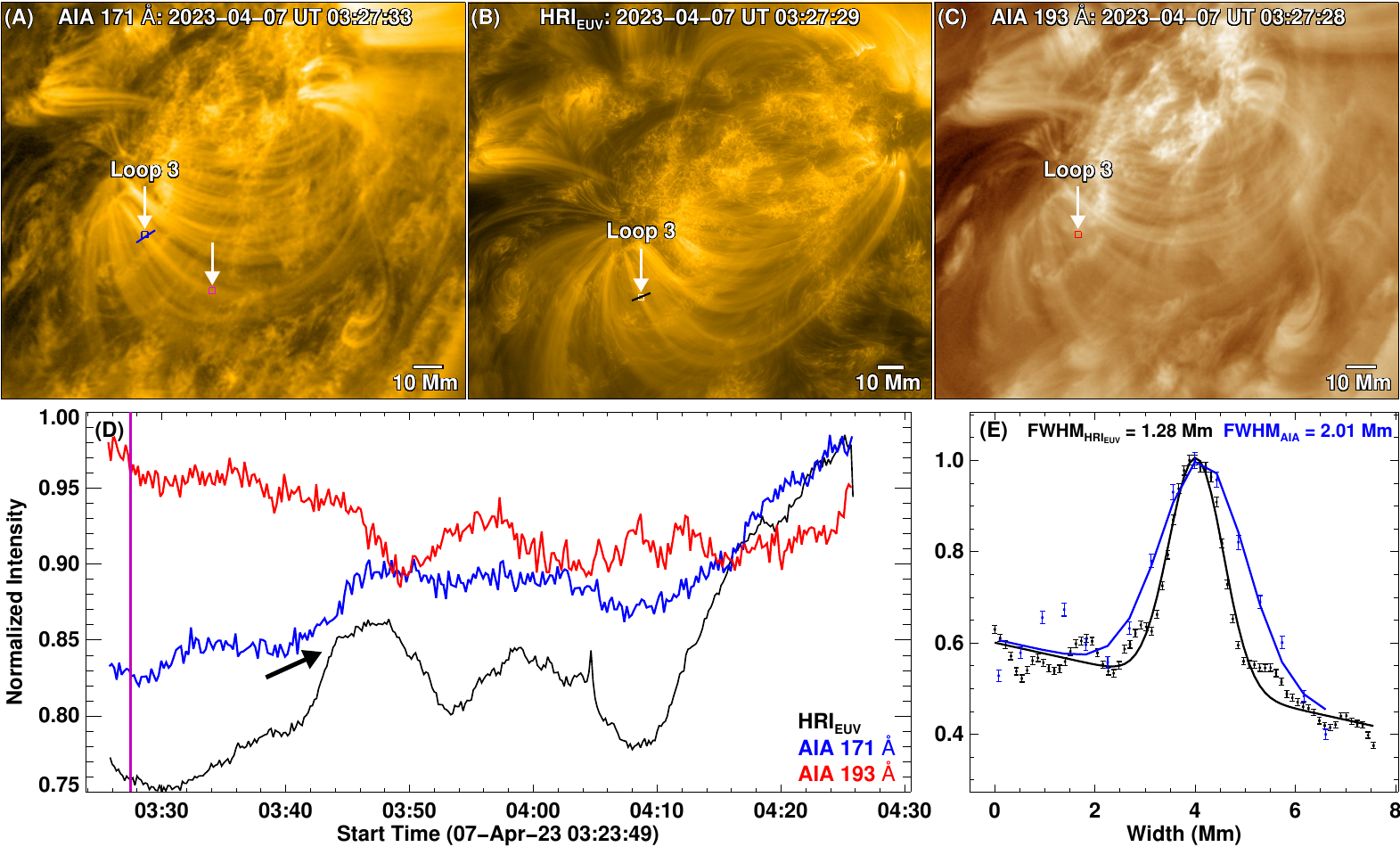}
 \caption{Evolution of loop 3. The panel layout, color, and size of the boxes for the light curves, and the slit widths are similar to those in Fig. \ref{loop1}. The average intensity within the 5~$\times$~5 pixel magenta box in panel (A) is used to derive the correspondingly colored light curve in panel (B) of Fig. \ref{loop_evolution}. The black arrow points to the intensity enhancement in the \hri\ light curve. An animation of panels (A)-(D) is available \href{https://www.dropbox.com/scl/fi/bm5fa0fz6w3dt8glzy7pw/23.mp4?rlkey=u2ekii9ucdhj83lsyvtubcg8y&dl=0}{online}.}
 \label{loop3}
\end{figure*}

We also find a case where the loop shows a similar overarching trend in both \hri\ and AIA 171~{\AA} filter but notable dissimilarities emerge within the evolutionary pattern. Figure \ref{loop3} displays the evolution of loop 3 close to its footpoint. Light curves in Panel (D) showcase that the overall normalized intensity is increasing throughout the observational period in both instruments. However, two prominent intensity peaks in the \hri\ light curves (the left peak is highlighted by a black arrow) are absent in both AIA filters. The presence of a bright moss region in the background is a potential contributor to these discrepancies. This background emission, integrated along the LOS, could potentially influence the observed intensity.

\section{Influence of background emission}
\label{BG_influence_appendix}

\begin{figure*}
 \centering
 \includegraphics[width=.8\linewidth]{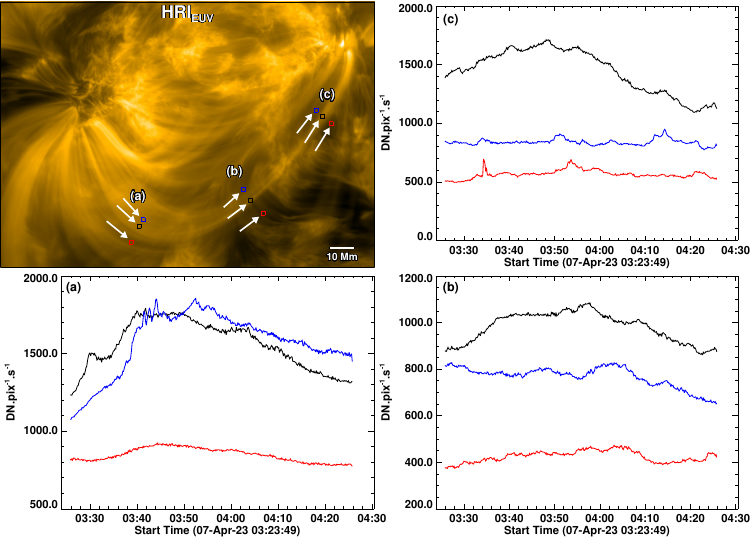}
 \caption{Influence of the background emission. The top left panel shows a one-hour time average of \hri\ images, starting at 03:23:49 UT on 07 April 2023. In the top left panel, (a), (b), and (c) denote the three loop segments for which the light curves are plotted in panels (a) - (c) respectively. White arrows indicate the color-coded boxes for which light curves are plotted in the subsequent panels.}
 \label{bg_influence}
\end{figure*}

To analyze the influence of background emission on the light curves from the loops or strands, we examine the intensity variations in regions adjacent to loop 3, as illustrated in Fig. \ref{bg_influence}. We position additional boxes on the northern and southern sides of those boxes used for plotting the light curves in Fig. \ref{loop_evolution}. In these new light curves, absolute intensity is plotted instead of normalized intensity. Upon examination, the blue box in segment (a) of this loop displays an evolution very similar to the box on the loop axis (black box). This similarity may be attributed to the blue box encompassing other strands of the same loop. In contrast, the red box in this segment corresponds to a low-background region and exhibits significantly lower brightness. However, it does show intensity patterns that are somewhat similar to those of the blue and black boxes, albeit at lower intensities. Light curves generated for segments (b) and (c), in comparison to their surrounding background emission, exhibit dissimilar behavior. Therefore, our analysis suggests that the integration of background emission in the loop segments, particularly where the background is not very bright, does not significantly influence the resulting light curves.

\FloatBarrier 

\section{Loop widths}

\begin{table}[h]
\caption{Comparison of Loop Widths}

\label{table_1}
\begin{tabular}{|c|c|c|}
\hline
Loop No. & Width in AIA 171 {\AA} (Mm) & Width in \hri~ (Mm) \\ \hline
1 & 1.42$~\pm~$0.07 & 1.88$~\pm~$0.02 \\ \hline
2 & 1.90$~\pm~$0.10 & 2.34$~\pm~$0.06 \\ \hline
3 & 1.99$~\pm~$0.05 & 1.28$~\pm~$0.01 \\ \hline
4 & 1.33$~\pm~$0.06 & 1.23$~\pm~$0.03 \\ \hline
5 & 2.46$~\pm~$0.02 & 3.76$~\pm~$0.01 \\ \hline
6 & 1.17$~\pm~$0.09 & 1.80$~\pm~$0.07 \\ \hline
7 & 1.66$~\pm~$0.10 & 1.04$~\pm~$0.03 \\ \hline
8 & 2.97$~\pm~$2.23 & 0.66$~\pm~$0.02 \\ \hline
9 & 0.56$~\pm~$0.15 & 0.69$~\pm~$0.04 \\ \hline
\end{tabular}
\tablefoot{Width measurements and their respective 1~$\sigma$ errors for the loops shown in Fig. \ref{loop_widths}. The uncertainties are generally an order of magnitude smaller than the measured widths.}
\end{table}

\end{appendix}

\end{document}